\documentclass[preprint]{revtex4}
\usepackage[dvips]{graphicx}
\usepackage{color}

\usepackage{graphicx}
\usepackage{amssymb}
\usepackage{amstext}

\providecommand{\red}[1]{\textcolor{red}{#1}}

\begin{document}

\vspace*{2em}
\begin{center}
{\bf \Large {\red{}} Growth-dependent bacterial susceptibility to ribosome-targeting antibiotics}
\end{center}
\vspace{2em}

\begin{center}
{\large Philip Greulich$^{1,2}$, Matthew Scott$^3$, Martin R. Evans$^2$\\ and  Rosalind J. Allen$^2$}\\[1em]

\end{center}

\noindent {\large \em  1: Cavendish Laboratory, University of Cambridge, J J Thompson 
Avenue, Cambridge 
CB3 0HE, United Kingdom}\\
{\large \em  2: SUPA, School of Physics and Astronomy, University of Edinburgh, James Clerk Maxwell Building, Peter Guthrie Tait Road, Edinburgh EH9 3FD, United Kingdom}\\
{\large \em  3: Department 
of Applied Mathematics, University of Waterloo, Waterloo, Ontario N2L 3G1 Canada}\\

\vspace{4cm}\noindent {\large \em Corresponding author: Rosalind J. Allen, rallen2@staffmail.ed.ac.uk}

\vspace*{3em}

\newpage

\begin{center}
{\bf Abstract}\\
\end{center}
\noindent Bacterial growth environment strongly influences the efficacy of antibiotic treatment, with slow growth often being associated with decreased susceptibility. Yet in many cases the connection between antibiotic susceptibility and pathogen physiology remains unclear. We show that for ribosome-targeting antibiotics acting on \emph{Escherichia coli}, a complex interplay exists between physiology and antibiotic action; for some antibiotics within this class faster growth indeed increases susceptibility, but for other antibiotics the opposite is true. Remarkably,  these observations can be explained by a simple mathematical model that combines drug transport and binding with physiological constraints. Our model reveals that growth-dependent susceptibility is controlled by a single parameter characterizing the `reversibility' of antibiotic transport and binding. This parameter provides a 
spectrum-classification of antibiotic growth-dependent efficacy that appears to correspond at its extremes to existing binary classification schemes. In these limits the model predicts universal, parameter-free limiting 
forms for growth inhibition curves. The model also leads to non-trivial predictions for the drug susceptibility of a translation-mutant strain of \emph{E. coli}, which we verify experimentally.  Drug action and bacterial metabolism are mechanistically complex; nevertheless this study illustrates how coarse-grained models can 
be used to integrate pathogen physiology into drug design and treatment strategies.

\newpage

\begin{center}
{\bf Introduction}
\end{center}
\vspace*{1ex}

{Q}uantitative predictions for the inhibition of bacterial growth by antibiotics are essential for 
the design of treatment strategies~\cite{Bacterial_treatment} and for controlling the evolution 
of antibiotic resistance \cite{antibiotic_gradients, 
gradients_hermsen,antibiotic_evolution,deris}. 
The efficacy of antibiotic treatment can be 
strongly affected by changes in pathogen physiology, such as  biofilm 
formation~\cite{Biofilm_review}, switching to persister states~\cite{lewis_review}, and 
responses to  metabolic stimuli~\cite{Collins_persister}; with slow bacterial growth often being associated with decreased antibiotic susceptibility  \cite{cozens,tuomanen,millar}. Yet, despite its importance, in most cases the connection 
between bacterial physiology and antibiotic susceptibility remains unclear. Here  we show that 
for ribosome-targeting antibiotics in \emph{Escherichia coli} a strong correlation exists between physiology, controlled by 
the nutrient quality of the growth environment, and antibiotic susceptibility. 

Ribosome-targeting antibiotics constitute a major class of antibacterial drugs in current 
clinical use. Within this class, different drugs bind to different ribosomal target sites, 
inhibit different aspects of ribosome function and  may bind to their target with varying degrees of reversibility~\cite{yonath_antibiotics,poehlsgaard}. We investigate four different ribosome-targeting antibiotics, two of which bind almost irreversibly, and two of which bind reversibly. Specifically, streptomycin and kanamycin are aminoglycosides  which bind irreversibly to the 30S ribosomal complex, inhibiting initiation and inducing mistranslation \cite{review_streptom_irr}. We also study the reversibly-binding drugs  tetracycline, which targets the 30S complex, inhibiting the binding of aminoacyl tRNA \cite{TET_binding}, and chloramphenicol, which targets the 50S ribosomal complex, preventing peptide bond formation \cite{Nierhaus1973,harvey_koch1980}.
 We find that the efficacies of these antibiotics exhibit qualitatively different responses to changes in the bacterial growth environment.

It has long been known that the ribosome content of a bacterial cell  correlates closely with 
its growth rate under conditions of exponential growth \cite{bremer_dennis,maaloe}. Recently, it 
has been shown that this phenomenon can be understood as a growth-rate dependent partitioning of 
the cell's translational resources between production of new ribosomes and production of other 
proteins \cite{scott_science,Hwa_CR}. This partitioning can be described by a set of 
empirically-determined constraints, analogous to the rules that govern the behaviour of electric 
circuits  \cite{scott_science,Scott2011}. Empirical growth constraints provide a physiological 
chassis into which mechanistic models for the expression of synthetic gene circuits, or endogenous genes, 
have been integrated \cite{Klumpp_Cell,klumpp2}. 

The fact that the cell's ribosome content is growth-rate dependent suggests that the efficacy of ribosome-targeting antibiotics should likewise exhibit growth-rate dependence. We demonstrate that 
bacterial susceptibility to ribosome-targeting antibiotics does indeed depend strongly on the 
nutrient environment as characterized by the bacterial growth rate prior to antibiotic treatment. Surprisingly, although the four antibiotics used in our study share the same target,  
we observe contrasting forms for the efficacy-growth rate relations of different antibiotics. 

These intriguing results can be explained by a simple mathematical model for antibiotic 
transport and ribosome binding which incorporates the empirical growth constraints; growth inhibition relations which are predicted by the model are in quantitative agreement with our data for both wild type and mutant strains of {\em E. coli}. A single dimensionless parameter, which characterizes the reversibility of transport and binding relative to the drug-free growth rate, emerges from our analysis, providing a simple way to predict how changes in  antibiotic chemistry, pathogen genetics or physiological state will affect drug response. This `reversibility parameter' provides a robust classification of ribosome-targeting 
antibiotics according to their  growth-rate efficacy relations, with implications for clinical 
practice and for the evolution of antibiotic resistance. In particular, reversible antibiotics are predicted to work better on fast-growing infections, whereas irreversible antibiotics are more effective for slow-growing pathogens. From a wider perspective, the approach taken here, in which empirical physiological constraints are coupled with models for molecular mode-of-action, could reveal similar surprising growth-rate efficacy relations in other classes of antibiotics.

\begin{figure}
\begin{center}
\includegraphics[width=0.5\textwidth]{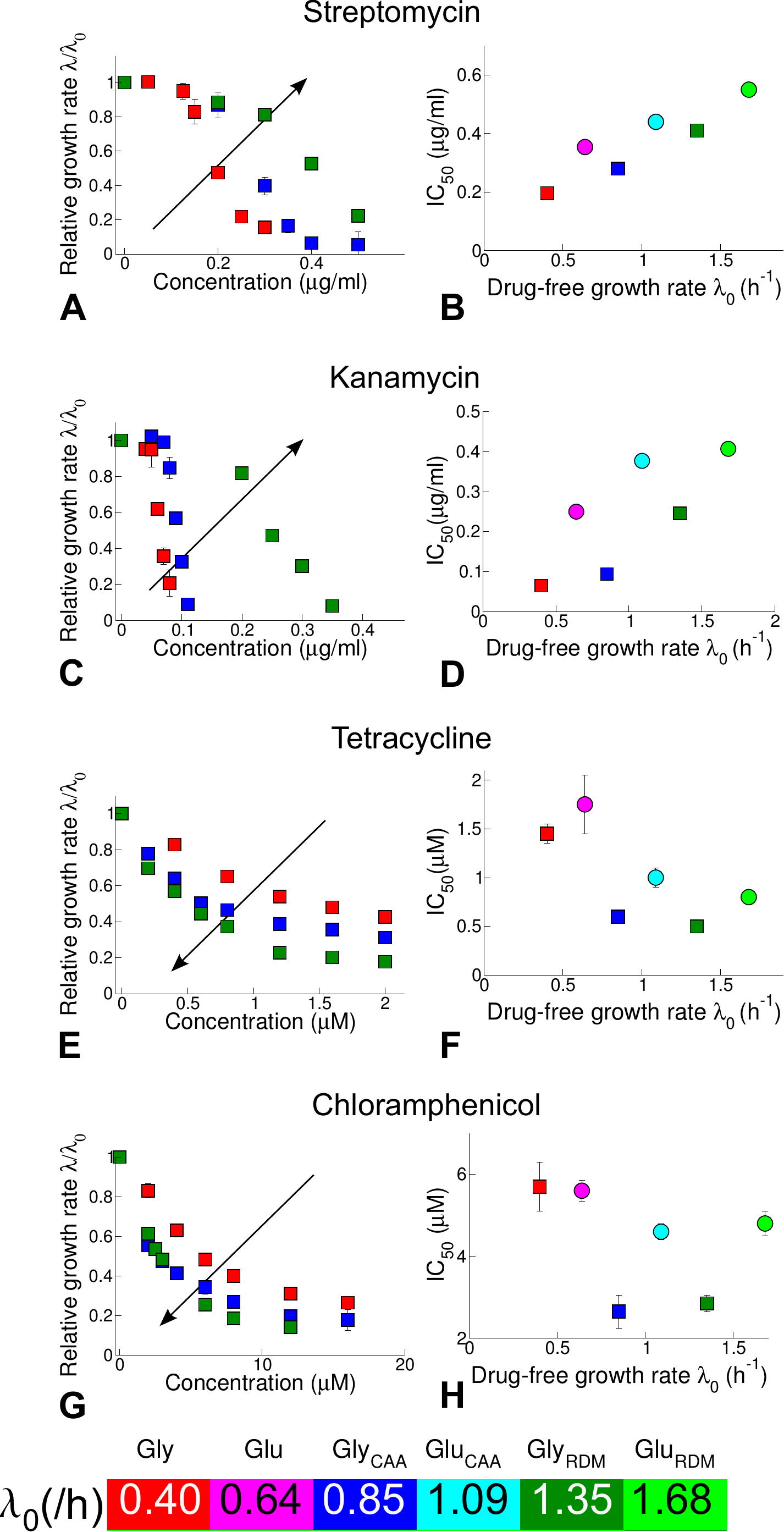} 
\end{center}
\caption{Antibiotic susceptibility depends on nutrient quality for four ribosome-targeting 
antibiotics: irreversibly-binding antibiotics streptomycin ({\bf A} \& {\bf B}) and kanamycin ({\bf C} 
\& {\bf D}), and reversibly-binding antibiotics tetracycline ({\bf E} \& {\bf F}) and  
chloramphenicol ({\bf G} \& {\bf H}). The left panels show the growth rate $\lambda$ of  
{\em{E. coli}} MG1655 relative to the drug-free growth rate $\lambda_0$, as a function of the 
antibiotic concentration.  Growth inhibition data are shown for media with glycerol as the 
carbon source. The arrows indicate increasing drug-free growth rate $\lambda_0$. The right 
panels show the half-inhibition concentration IC$_{50}$ as a function of the drug-free growth 
rate $\lambda_0$. Carbon sources are denoted by symbol: glucose (circles), glycerol (squares), 
and error bars denote the standard deviation among repeated measurements (Tables S2 and S3). Media are variants of Neidhardt's MOPS buffered medium \cite{Neidhardt_Media}; see Methods for details.}

\end{figure}

\newpage

\begin{center}
{\bf Results}
\end{center}

\noindent {\bf{Antibiotic efficacy depends on growth rate}}
\vspace*{1ex}

\noindent To investigate the link between bacterial growth environment and susceptibility to 
ribosome-targeting antibiotics,  we measured growth inhibition curves (exponential growth rate 
as a function of antibiotic concentration) for  {\em{E. coli}} cells  on  media of increasing 
nutrient quality. As the nutrient quality increases, so too does the `drug-free growth rate'  
$\lambda_0$, {\em{i.e.}} the exponential growth rate in the absence of antibiotic (Fig.~1, colorbar and 
Table S1). For the four ribosome-targeting antibiotics streptomycin, kanamycin, tetracycline and 
chloramphenicol, the growth inhibition curves indeed exhibit a strong dependence on the 
drug-free growth rate $\lambda_0$ (Fig.~1, left panels and Table S2). 

Bacterial susceptibility to antibiotic can be quantified by the IC$_{50}$: the antibiotic 
concentration needed to halve the bacterial growth rate.  Plotting the IC$_{50}$ as a function 
of the drug-free growth rate $\lambda_0$, we observe contrasting trends between antibiotics 
(Fig.~1, right panels and Table S3). For the irreversibly-binding antibiotics streptomycin and 
kanamycin, the  IC$_{50}$ increases with nutrient quality, {\em{i.e.}} faster-growing cells are 
less susceptible to antibiotic. In  contrast, for  the reversibly-binding antibiotics tetracycline 
and chloramphenicol, the IC$_{50}$ predominantly decreases as nutrient quality increases, 
{\em{i.e.}} faster-growing cells are more susceptible to antibiotic treatment. Data sets for 
glycerol and glucose-based media show distinct trends in IC$_{50}$ with drug-free growth rate.  
The shapes of the growth inhibition curves  also differ markedly between the two groups of antibiotics: we 
observe threshold-like  inhibition, {\em i.e.} a sharp decrease in growth rate,  for  streptomycin and kanamycin 
(Fig.~1A \& C), and more gradual inhibition  for  tetracycline and 
chloramphenicol (Fig.~1E \& G). Despite having similar targets, these antibiotics appear to 
respond to changes in cell physiology in very different ways.\\

\noindent {\bf Mathematical model}
\vspace*{1ex}

\noindent Our experimental data can be explained by a simple mathematical model. In our model, antibiotic 
molecules enter a bacterial cell and bind to ribosomes, while at the same time, new ribosomes 
are synthesized and the cell contents are diluted by growth. Our model is placed within a 
physiological context via the empirical growth constraints
~\cite{scott_science,Scott2011}.

 \begin{figure}
\begin{center}
\includegraphics[width=0.6\textwidth]{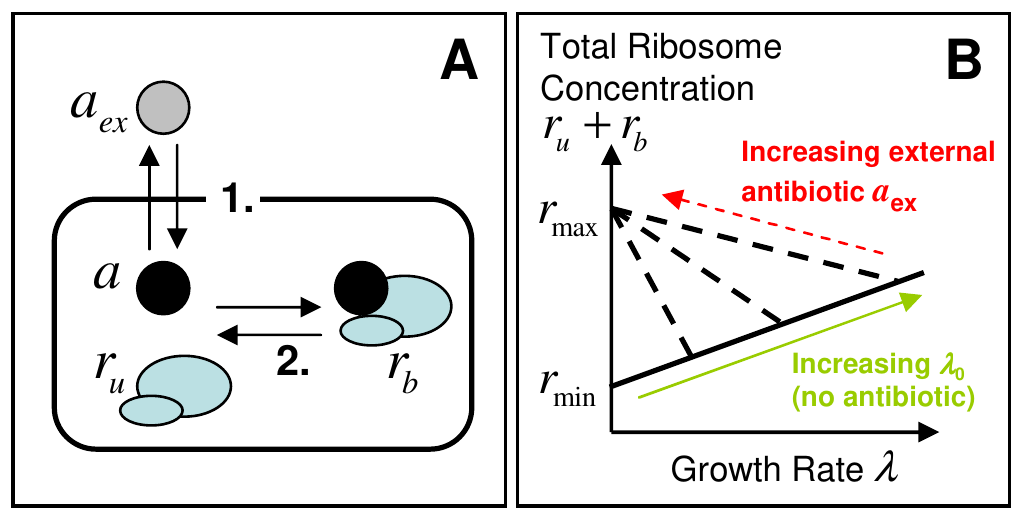}
\end{center}\label{Fig_schematic1} 
\caption{Schematic view of the model and its dynamics. {\bf A.}  The model is focused on three 
state variables: the intracellular concentration of antibiotic $a$, the concentration 
$r_{\mathrm{u}}$ of ribosomes unbound by antibiotic and the concentration $r_{\mathrm{b}}$ of 
antibiotic-bound ribosomes. Two mechanisms drive the dynamics: 1. \emph{Transport} across the 
cell membrane and 2. \emph{Binding} of ribosomes and antibiotic.  {\bf B.} Constraints arising 
from empirical relations between ribosome content and growth rate. Scott {\it et al.} 
{\cite{scott_science}} measured total ribosome content as a function of growth rate. When growth 
rate is varied by nutrient composition, in the absence of antibiotics, ribosome content 
$r_{\mathrm{u}}$ correlates positively with growth rate $\lambda$, increasing linearly from a 
minimum concentration of inactive ribosomes $r_\mathrm{min}$ (solid line). When growth rate is 
decreased by imposing translational inhibition, total ribosome content 
$r_\mathrm{tot}=r_\mathrm{u}+r_\mathrm{b}$ increases, reaching  a maximum  $r_\mathrm{max}$ as 
growth rate decreases to zero (dashed lines). Note that Scott {\em{et al.}} measured ribosome 
mass fraction; here we translate these to concentrations (see Supplementary Information, Fig. S1).  
\\}
\end{figure}

In the model,  the state of the cell is described by the intracellular concentration of 
antibiotic $a$, the concentration $r_{\mathrm{u}}$ of ribosomes unbound by antibiotic and the 
concentration $r_{\mathrm{b}}$ of antibiotic-bound ribosomes (Fig.~2A). Two 
mechanisms drive the dynamics: 1. transport of extracellular antibiotic $a_{\mathrm{ex}}$ into 
the cell at rate $J(a_{\mathrm{ex}},a)=P_{\mathrm{in}} a_{\mathrm{ex}} - P_{\mathrm{out}} a$, 
where $P_{\mathrm{in}}$ and  $P_{\mathrm{out}}$ quantify the permeability of the cell membrane 
in the inward and outward directions; and 2. binding of ribosomes and antibiotic 
$f(r_{\mathrm{u}},r_{\mathrm{b}},a)=- k_{\rm on}\,a\, (r_{\mathrm{u}} - r_{\mathrm{min}}) + 
k_{\rm off}\, r_{\mathrm{b}} $, with binding and unbinding rate constants $k_{\rm on}$ 
and $k_{\rm off}$, respectively, and equilibrium dissociation constant $K_{\rm D} = k_{\rm off}/k_{\rm on}$ (the 
inactive fraction $r_\mathrm{min}$ is assumed not to bind the antibiotic). In exponential 
growth, cell contents are diluted at rate $\lambda$, new ribosomes are synthesized at rate 
$s(\lambda)$, and the dynamics of the system are governed by the following equations: 
\begin{eqnarray}
\label{rate_eq_simple1}
\frac{da}{dt} &=& - \lambda \, a + f(r_{\mathrm{u}},r_{\mathrm{b}},a) + J(a_{\mathrm{ex}},a), \\  
\label{rate_eq_simple2}
\frac{dr_{\mathrm{u}}}{dt} &=& -\lambda \, r_{\mathrm{u}} + f(r_{\mathrm{u}},r_{\mathrm{b}},a)  
+ s(\lambda), \\  \label{rate_eq_simple3}
\frac{dr_{\mathrm{b}}}{dt}  &=& - \lambda \, r_{\mathrm{b}} -f(r_{\mathrm{u}},r_{\mathrm{b}},a).
\end{eqnarray}

This model is coupled to cell physiology via the empirical relations of Scott {\em et al} {\cite{scott_science}}, which link the growth rate 
$\lambda$ and ribosome synthesis rate $s(\lambda)$ to the ribosome concentration; 
these act as constraints on the dynamical equations \ref{rate_eq_simple1}-\ref{rate_eq_simple3}. 
The first empirical growth constraint states that the unbound ribosome content $r_{\rm u}$ and the growth rate 
$\lambda$ are linearly proportional:
\begin{eqnarray}
\label{gl1}
r_{\mathrm{u}}&=& \lambda/\kappa_{\rm t} +r_{\mathrm{min}}.
\end{eqnarray}
Here, $r_\mathrm{min}=19.3 \mu$M is the minimal unbound ribosome content needed for growth and 
the translational capacity $\kappa_{\rm t}=0.06 \mu$M$^{-1}$h$^{-1}$ is related to the maximum 
peptide elongation rate~\cite{Klumpp2013}. This relation emerges from experiments in which the 
growth rate is varied by changing the nutrient source in the absence of antibiotic (green arrow 
in Fig.~2B; see also the Supplementary Information). The second empirical growth constraint  
describes how the ribosome content is upregulated in response to translational inhibition. Upon 
decreasing the growth rate by translational inhibition (for a fixed nutrient source), the total 
ribosome content $r_{\mathrm{tot}}$ increases linearly, reaching a fixed maximal value $r_\mathrm{max}=65.8 \mu$M 
as $\lambda \to 0$~\cite{scott_science} (red arrow in Fig.~2B; see also the 
Supplementary Information). This can be expressed mathematically as
\begin{eqnarray}
\label{gl2}
r_{\mathrm{tot}}&=&r_{\mathrm{u}}+r_{\mathrm{b}}=r_\mathrm{max}-\lambda \Delta r 
\left({\frac{1}{\lambda_0}-\frac{1}{\kappa_{\rm t} \Delta r}}\right),
\end{eqnarray}
where $\Delta r = r_\mathrm{max}- r_\mathrm{min} = 46.5 \mu$M is the dynamic range of the 
ribosome concentration. The implication of the second empirical growth constraint, Eq. \ref{gl2}, is that cells 
that are initially growing more slowly have a greater capacity to upregulate their ribosome 
content upon antibiotic challenge (steeper slope of the dashed line in 
Fig.~2B) than those that are initially growing fast; {\em{i.e.}}, slowly growing cells can increase their ribosome content with little resulting change in their growth rate.   Adding together Eqs 2 and 3 at steady state ($dr_{\rm u}/dt = dr_{\rm b}/dt = 0$) shows that the ribosome synthesis rate $s(\lambda)$ is the product of growth rate and 
total ribosome content, 
\begin{eqnarray}
\label{synth_rate}
s(\lambda) = \lambda r_{\rm tot}=\lambda 
\left[r_\mathrm{max}-\lambda \Delta r \left({\frac{1}{\lambda_0}-
\frac{1}{\kappa_{\rm t} \Delta r}}\right)\right].
\end{eqnarray}\\

\noindent {\bf Quantitative results for growth-inhibition curves}

\noindent Solving the model equations 1-3 at steady state, together with the  physiological constraints, Eqs 4 and 5, produces a universal equation that links the steady-state relative growth rate $\lambda/\lambda_0$ to the extracellular antibiotic concentration 
$a_{\rm ex}$ (see Supplementary Information; here we have assumed that  the antibiotic binding rate $k_{\rm on}$ typically exceeds the translational capacity $\kappa_{\bf t}$ by several orders of magnitude, $k_{\rm on}\gg \kappa_{\bf t}$). This equation is
\begin{eqnarray}\label{eq:S14}
 0 &=&   \left( {\frac{\lambda }
{{\lambda _0 }}} \right)^3
 - \left( {\frac{\lambda }
{{\lambda _0 }}} \right)^2  + \left( {\frac{\lambda }
{{\lambda _0 }}} \right)\left[\frac{1}{4}\left(\frac{\lambda_0^{\rm{*}}}{\lambda_0}\right)^2 + 
\frac{a_{\rm ex}}
{2\,{\rm IC}_{50}^*}\left(\frac{\lambda_0^{\rm{*}}}{\lambda_0}\right) \right] - 
\frac{1}{4}\left(\frac{\lambda_0^{\rm{*}}}{\lambda_0}\right)^2.
\end{eqnarray}
Remarkably, Eq. 7 states that the growth-dependent antibiotic susceptibility is controlled by only two parameter combinations, defined as follows: a rate $\lambda_0^{\rm{*}}$, which characterizes the reversibility of antibiotic transport and binding:
\begin{equation}
\label{eq:lamStar}
\lambda_0^{\rm{*}}= 2\sqrt{P_{\rm out}\kappa_t K_D},
\end{equation}
and  a concentration scale
\begin{equation}
 {\rm IC}_{50}^* =\frac{\Delta r \lambda_0^{\rm{*}}}{2P_{\rm in}}.
\end{equation}
In the model, $\lambda_0^{\rm{*}}$ is used to normalize the drug-free growth rate $\lambda_0$ and ${\rm IC}_{50}^*$ is used to normalize the extracellular antibiotic concentration $a_{\rm ex}$, and later the half-inhibition concentration IC$_{50}$.

Predictions for growth inhibition curves can be obtained by solving Eq.~7; the shapes of these curves depend only on the value of $\lambda_0^{\rm{*}}$. For small values of 
$\lambda_0^{\rm{*}}$ (the irreversible limit), the model predicts a  discontinuous drop in growth rate at the IC$_{50}$, as we see in our data for streptomycin and kanamycin (Fig. 1 and Fig.~3). Interestingly, in this case the model predicts 
 a bistable dependence of growth rate on antibiotic concentration at the level of individual 
cells (Fig. S2). For larger values of $\lambda_0^{\rm{*}}$ (the reversible limit), the model instead predicts a smooth 
decrease in growth rate over a wide range of antibiotic concentrations, as we observe for tetracycline and chloramphenicol  (Fig. S2, Fig. 1  and Fig. 3). 

Fitting the model to the data via the parameters $\lambda_0^*$ and IC$_{50}^*$ yields excellent agreement for tetracycline and  chloramphenicol, and reasonable agreement for streptomycin and kanamycin (Fig. 3). In all cases, the fitted parameters $\lambda_0^{\rm{*}}$ and ${\rm IC}_{50}^*$ differ between  the two carbon sources (Table S3), suggesting carbon-source effects on  transporter-mediated influx and outflux  ($P_{\rm in}$ and $P_{\rm out}$, respectively)~\cite{Collins_persister}. The fitted parameters are in good agreement with biochemical parameter values available from literature data (Table S4) and are consistent with the fact that aminoglycosides are believed to bind and be transported irreversibly (small $\lambda_0^*$)~ \cite{review_streptom_irr}, whereas for tetracycline and chloramphenicol both transport and binding processes are reversible ($\lambda_0^*$) ~\cite{TET_review_berens,harvey_koch1980}. For kanamycin the agreement is less accurate than for streptomycin; this deviation may be ascribed to the presence of an additional binding site on the ribosome with comparable affinity~\cite{biochem_antimicrobials_book,misumi}, a property not considered in our model. Nonetheless, our model correctly predicts the sigmoidal form of kanamycin's growth-inhibition curve and the decreasing susceptibility with growth rate. \\

\begin{figure}[t!] 
\begin{center}
\includegraphics[width=0.5\textwidth]{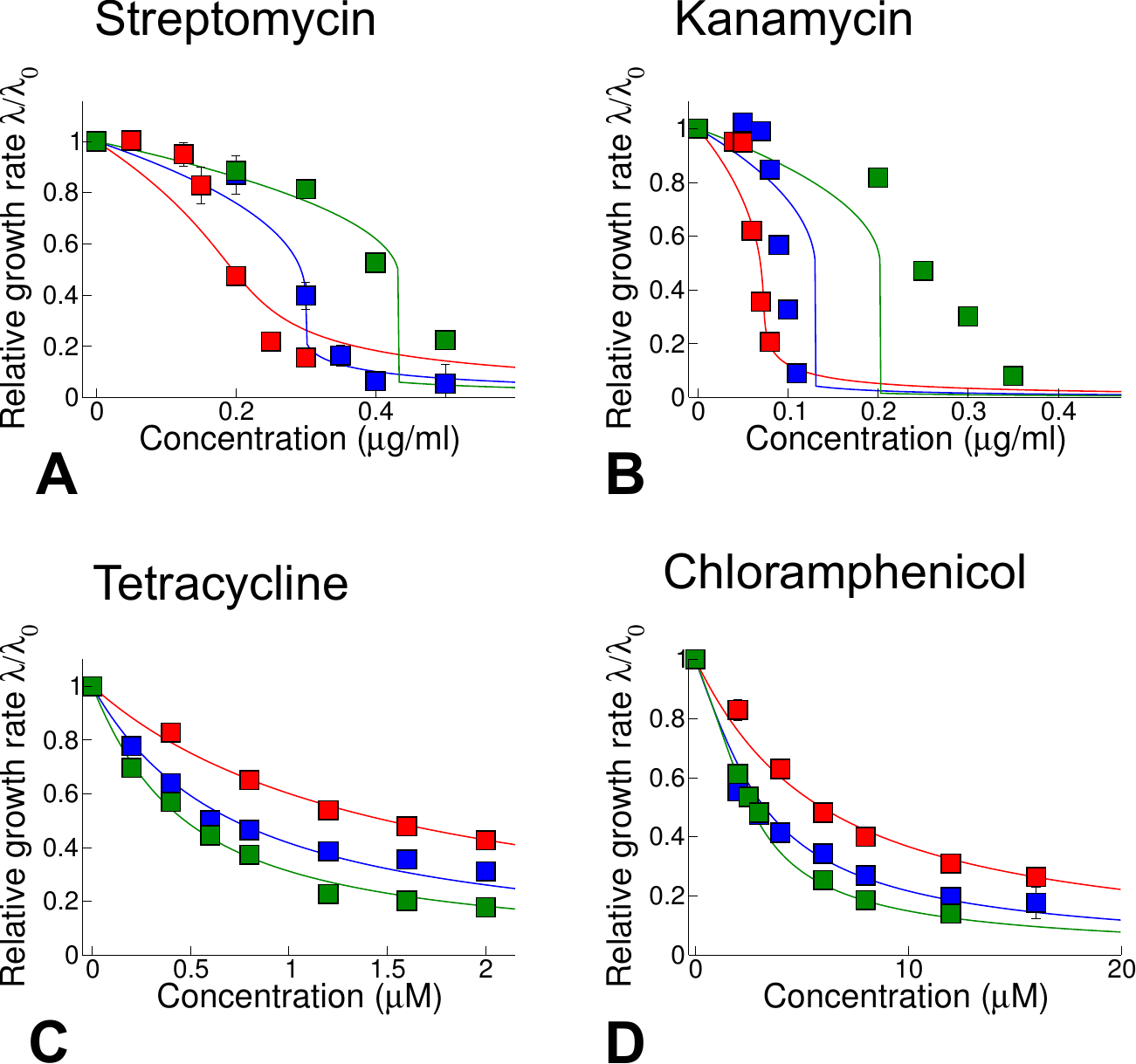} 
\end{center}
\caption{Model fits to growth inhibition curve data. The parameters $\lambda_0^{\rm{*}}$ and  
${\rm IC}_{50}^*$ are obtained by numerical fitting of the solution of the cubic equation, Eq. 
7, to our experimental growth inhibition curves. Data sets for different drug-free growth rates ({\em{i.e.}} the different curves in each panel) were fitted simultaneously with the same values of  $\lambda_0^{\rm{*}}$ and  
${\rm IC}_{50}^*$, but separate fits were obtained for glycerol-based 
and glucose-based media. Here we show the resulting fits for  glycerol-based media (symbols as in Fig.~1.). The parameters obtained by this 
procedure are: Streptomycin, glycerol: $\lambda_0^{\rm{*}}=0.31$h$^{-1}$, 
${\rm IC}_{50}^*=0.19\mu$g/ml,  streptomycin, glucose: $\lambda_0^{\rm{*}}=0.57$h$^{-1}$, 
${\rm IC}_{50}^*=0.36\mu$g/ml, kanamycin, glycerol: $\lambda_0^{\rm{*}}=0.17$h$^{-1}$, 
${\rm IC}_{50}^*=0.05\mu$g/ml,  kanamycin, glucose: $\lambda_0^{\rm{*}}=0.47$h$^{-1}$, 
${\rm IC}_{50}^*=0.26\mu$g/ml, tetracycline, glycerol: $\lambda_0^{\rm{*}}=5.2$h$^{-1}$, 
${\rm IC}_{50}^*=0.23\mu$M,  tetracycline, glucose: $\lambda_0^{\rm{*}}=6.3$h$^{-1}$, 
${\rm IC}_{50}^*=0.36\mu$M, chloramphenicol, glycerol: $\lambda_0^{\rm{*}}=1.8$h$^{-1}$, 
${\rm IC}_{50}^*=2.5\mu$M,  chloramphenicol, glucose: $\lambda_0^{\rm{*}}=1.5$h$^{-1}$, 
${\rm IC}_{50}^*=4.1\mu$M. These values of $\lambda_0^*$ and ${\rm IC}_{50}^*$ are  compared to 
literature data in Table S4. Similar results are obtained if we instead fit our data directly to the predicted universal relation for IC$_{50}(\lambda_0)$ (Eq. 10); see Supplementary Information and Fig. S3.\\}
\end{figure}

\noindent {\bf Universal growth-dependent antibiotic susceptibility curve}

\noindent One of the major insights provided by the model is a simple explanation for the contrasting trends in   growth-dependent 
susceptibility for different antibiotics which we observe in our experiments. Substituting $a_{\rm ex}={\rm IC}_{50}$ and 
$\lambda=\lambda_0/2$ into Eq.~7, we find that, for all antibiotics, the growth-rate dependence of the half-inhibition concentration ${\rm IC}_{50}$  is predicted to fall onto a 
universal `growth-dependent susceptibility' curve
\begin{equation}
\frac{\mbox{IC}_{50}}{\mbox{IC}_{50}^{\rm *}} = 
\frac{1}{2}\;\left[{\frac{\lambda_0}{\lambda_0^{\rm{*}}}+
\frac{\lambda_0^{\rm{*}}}{\lambda_0}}\right].
\label{eq:universal}
\end{equation}
 Eq. 10 is derived in the Supplementary Information and holds for $k_{\rm on} \gg \kappa_t$. Rescaling our data using the values of  $\lambda_0^{\rm{*}}$ and 
IC$_{50}^{\rm *}$ obtained from the growth-inhibition curve fits of Fig.~3 (and the equivalent fit for the glucose-based 
media; Table S3), Figure~4 shows that our data indeed collapse on to this universal curve. 

If the drug-free growth rate $\lambda_0$ exceeds the critical reversibility rate $\lambda_0^{\rm{*}}$, the model (Eq. 10) predicts that the IC$_{50}$ will increase with $\lambda_0$; {\em{i.e.}} fast-growing cells will be less susceptible, as we observe for streptomycin and kanamycin. In contrast, if the drug-free growth rate $\lambda_0$ is less than the critical reversibility rate $\lambda_0^{\rm{*}}$, Eq. 10 predicts that the IC$_{50}$ will decrease as $\lambda_0$ increases; {\em{i.e.}} fast-growing cells will be more susceptible, as we observe for tetracycline and chloramphenicol. The critical parameter IC$_{50}^*$ provides a growth-rate independent scale for the extracellular antibiotic concentration;  we find that an antibiotic concentration $a_{\rm ex} > {\rm IC}_{50}^*$ is required for effective growth inhibition, regardless of the drug-free growth rate.  

The universal growth-dependent susceptibility curve, Eq. 10 (Fig. 4), suggests that the ratio $(\lambda_0/\lambda_0^{\rm *})$ of the drug-free growth rate $\lambda_0$ to the `reversibility' rate $\lambda_0^{\rm *}=\sqrt{\kappa_{\rm t} P_{\rm out} k_{\rm off}}$ provides a natural spectrum classification of antibiotic action, integrating growth environment (through $\lambda_0$) with antibiotic chemistry and pathogen genetics (through the molecular parameters which are combined in $\lambda_0^{\rm *}$). Drug-pathogen interactions characterized by small values of $\lambda_0^{\rm{*}}$ are predicted to behave like our 
irreversible antibiotics (streptomycin and kanamycin); showing decreased efficacy under rich nutrient conditions. Drug-pathogen interactions characterized by large values of $\lambda_0^{\rm{*}}$ are expected to behave like 
the reversible antibiotics in our study (chloramphenicol and tetracycline); showing increased efficacy under rich nutrient 
conditions. Drugs with 
values of  $\lambda_0^{\rm{*}}$ close to the drug-free growth rate $\lambda_0$ achievable in experiments may 
show non-monotonically varying susceptibility as nutrient quality is varied; our data suggest 
this may in fact be the case for chloramphenicol (Fig. 1H), in agreement with literature-value estimates for $\lambda_0^{\rm{*}}$ (Table S4).  Low outward permeability has been implicated in growth bistability and masking of resistance mutations~\cite{elf_antibiotics,fange_pnas} (see in particular the discussion of Ref.\cite{elf_antibiotics} in the Supplementary Text); we propose that irreversibility in 
binding and transport is a major determinant of growth-dependent antibiotic susceptibility.\\

\begin{figure}[t!] 
\begin{center}
\includegraphics[width=0.6\textwidth]{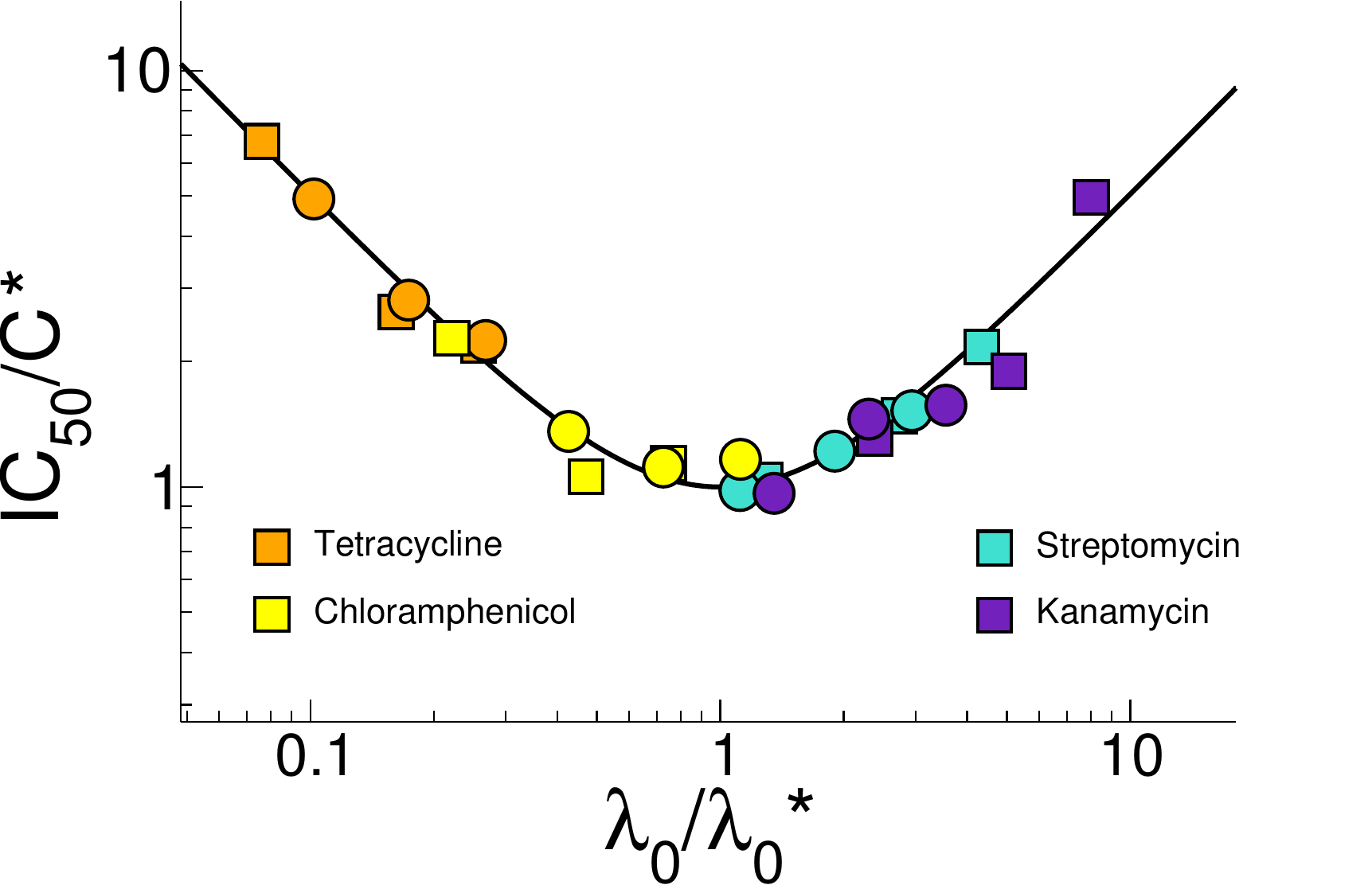} 
\end{center}
\caption{ Universal growth-dependent susceptibility curve. Data from the right panels of Fig. 1 
are rescaled by  $\lambda_0^{\rm{*}}$ and IC$_{50}^{\rm *}$, obtained by fitting our growth 
inhibition data (Fig.~3). The black line shows the model prediction for the universal curve, 
Eq.~10. \\}
\end{figure}

\noindent{\bf Simple predictions in the reversible and irreversible limits}

\noindent In the limiting cases of very large or very small $\lambda_0^{\rm{*}}$, {\em{i.e.}} the limits in which antibiotic transport and binding is either fully reversible or fully irreversible, the model leads to simple predictions for the growth inhibition curve and growth-rate dependence of the half-inhibition concentration 
IC$_{50}$. For small $\lambda_0^{\rm{*}}$ (the irreversible limit), a qualitatively different, 
discontinuous, form for the growth inhibition curve is predicted by Eq.~7: 
\begin{equation}
\frac{\lambda}{\lambda_0} = \frac{1}{2}
\left[1 + \sqrt{1 - \frac{a_{\rm ex}}{{\mathrm{IC}}_{50}}}\,\,\right],
\end{equation}
 for $a_{\rm ex} < \mathrm{IC}_{50}$ and zero for $a_{\rm ex} > \mathrm{IC}_{50}$. In this case,  
the  IC$_{50}={\rm IC}_{50}^*\lambda_0/(2\lambda_0^{\rm{*}}) = 
\lambda_0 \Delta r/(4P_{\mathrm{in}})$ increases linearly with the drug-free growth rate 
$\lambda_0$ (see Supplementary Information). For large $\lambda_0^{\rm{*}}$ (the reversible limit), the growth inhibition curve obtained from solving Eq.~7 is given by the smoothly-varying, Langmuir form:
\begin{equation}
 \frac{\lambda}{\lambda_0} = \frac{1}{1+a_{\mathrm{ex}}/{\rm{IC}}_{50}},
\end{equation}
 where $a_{\mathrm{ex}}$ is the extracellular antibiotic concentration and the  
IC$_{50}={\rm IC}_{50}^*\lambda_0^{\rm{*}}/(2\lambda_0) = 
K_D\times(P_{\mathrm{out}}/P_{\mathrm{in}})\times(\kappa_t \Delta r/\lambda_0)$ is  inversely 
proportional to the drug-free growth rate $\lambda_0$ (see Supplementary Information). 

 Scaling all our growth inhibition curves by the drug-free growth rate $\lambda_0$ and the 
half-inhibition concentration IC$_{50}$ we find that our combined data sets for the 
reversible and irreversible drugs collapse onto these two qualitatively distinct, 
parameter-free curves, as predicted by the model (Fig. 5) -- although, as expected, the quantitative agreement with the limiting-case theoretical prediction is not quite as good as with the full solution of the cubic equation (Fig. 3). \\

 \begin{figure}
\begin{center}
\includegraphics[width=0.8\textwidth]{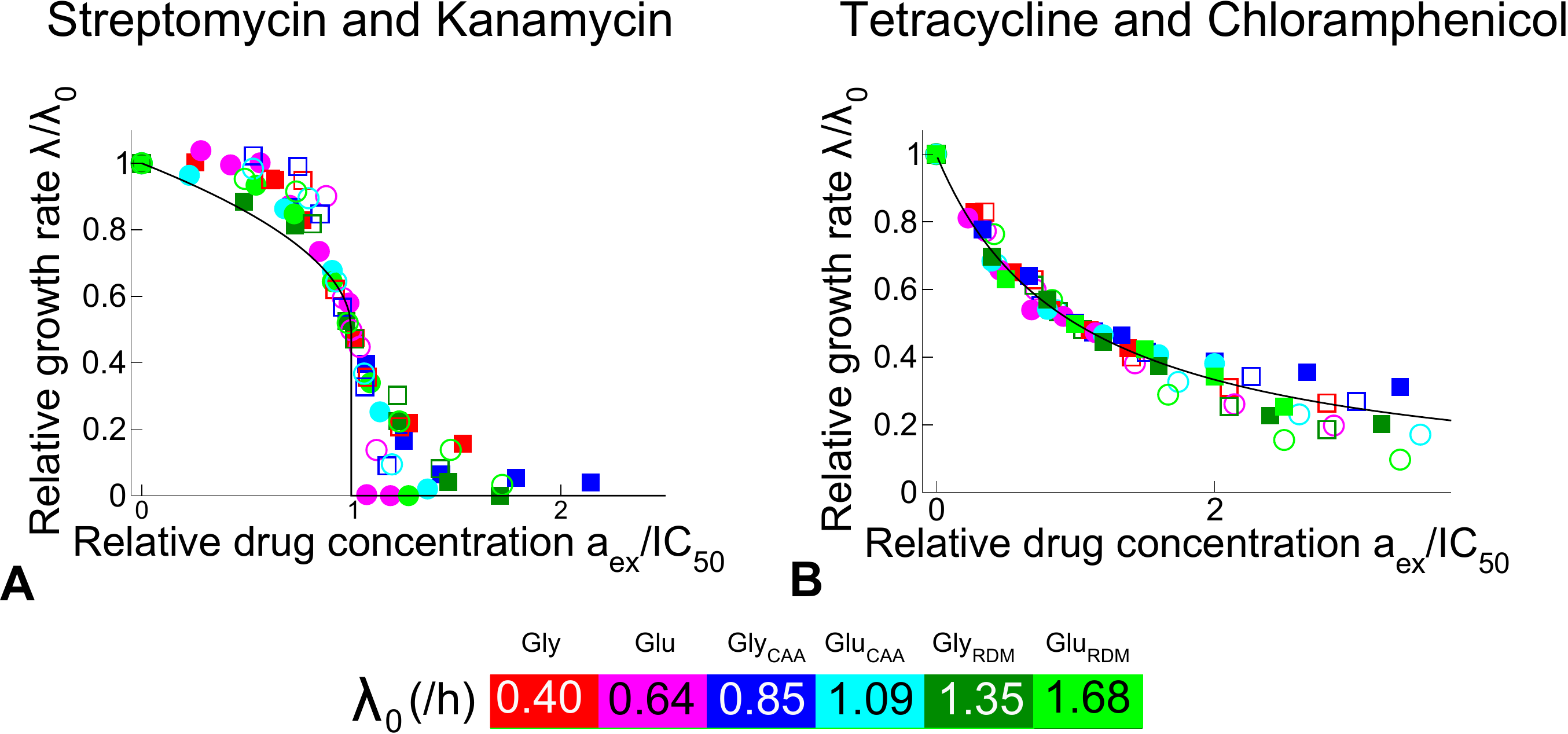}
\end{center}
\caption{Growth inhibition curves for our reversibly- and irreversibly-binding drugs collapse onto 
two qualitatively different limiting forms as predicted by the model. {\bf{A}}: Data for the 
irreversible antibiotics streptomycin (closed symbols) and kanamycin (open symbols) collapse 
onto $\lambda/\lambda_0 = (1/2)[1 + \sqrt{1 - a_{\rm ex}/{\mathrm{IC}}_{50}}]$ (black line). 
{\bf B}: Data for the reversible antibiotics tetracycline (closed symbols) and 
chloramphenicol (open symbols) collapse onto 
$\lambda/\lambda_0=1/[1+a_{\mathrm{ex}}/{\mathrm{IC}}_{50}]$ (black line).\\ }
\end{figure}

\noindent {\bf Testing the model predictions for a translation mutant strain of {\em E. coli}}
\vspace*{1ex}

In our model, the key parameters $\lambda_0^*$ and ${\mathrm{IC}}_{50}^*$ (defined in Eqs 8 and 9) depend on the translational capacity $\kappa_{\rm t}$. To test the predictions of the model, we used a strain of {\em E. coli} MG1655 in which the ribosome is mutated such that the peptide elongation rate is decreased~\cite{ruusala}, with a corresponding decrease in the translational capacity~\cite{scott_science}. Measuring the RNA-to-protein ratio, which is proportional to the ribosome concentration (\cite{scott_science}; see Supplementary Material) as a function of growth rate in the absence of antibiotics  and  using Eq.~4, we found that the translational capacity $\kappa_{\rm t}$ for the mutant is decreased by a factor of $0.65 $ relative to that of the wildtype, $\kappa_{\rm t}^{\rm MUT} = 0.65 \kappa_{\rm t}^{\rm WT} $ (Fig.~6A and Table S5). 

For the reversible antibiotic tetracycline, we expect that the ${\mathrm{IC}}_{50}$ is well-approximated by the limiting form, ${\mathrm{IC}}_{50}=(\kappa_t/\lambda_0) \times K_D \times (P_{\rm out}/P_{\rm in}) \times \Delta r$; thus the ratio of susceptibilities between the wildtype and mutant strains ${\mathrm{IC}}_{50}^{\rm WT}/{\mathrm{IC}}_{50}^{\rm MUT}$ should be proportional to the ratio of drug-free growth rates $\lambda_0^{\rm MUT}/\lambda_0^{\rm WT}$, with proportionality constant $\kappa_{\rm t}^{\rm WT}/\kappa_{\rm t}^{\rm MUT} = 1/0.65$. Indeed, when rescaled relative to the ${\mathrm{IC}}_{50}$ of the mutant in minimal media, our results for the wild-type ${\mathrm{IC}}_{50}$ values, measured for our 6 nutrient conditions, do fall on the predicted straight line with gradient 1/0.65 irrespective of carbon source (Fig. 6B; for raw data see Table S6). 

We also investigated the response of the translational mutant to the irreversibly-binding drug kanamycin. Here, the situation is more complex because the mutant confers partial resistance to kanamycin (and full resistance to streptomycin), meaning that other molecular parameters are likely to be altered along with $\kappa_{\rm t}$. Nevertheless, growth inhibition curves for the mutant in the presence of kanamycin are well-fitted by our model (Fig. S4).\\

 \begin{figure}
\begin{center}
\includegraphics[width=0.9\textwidth]{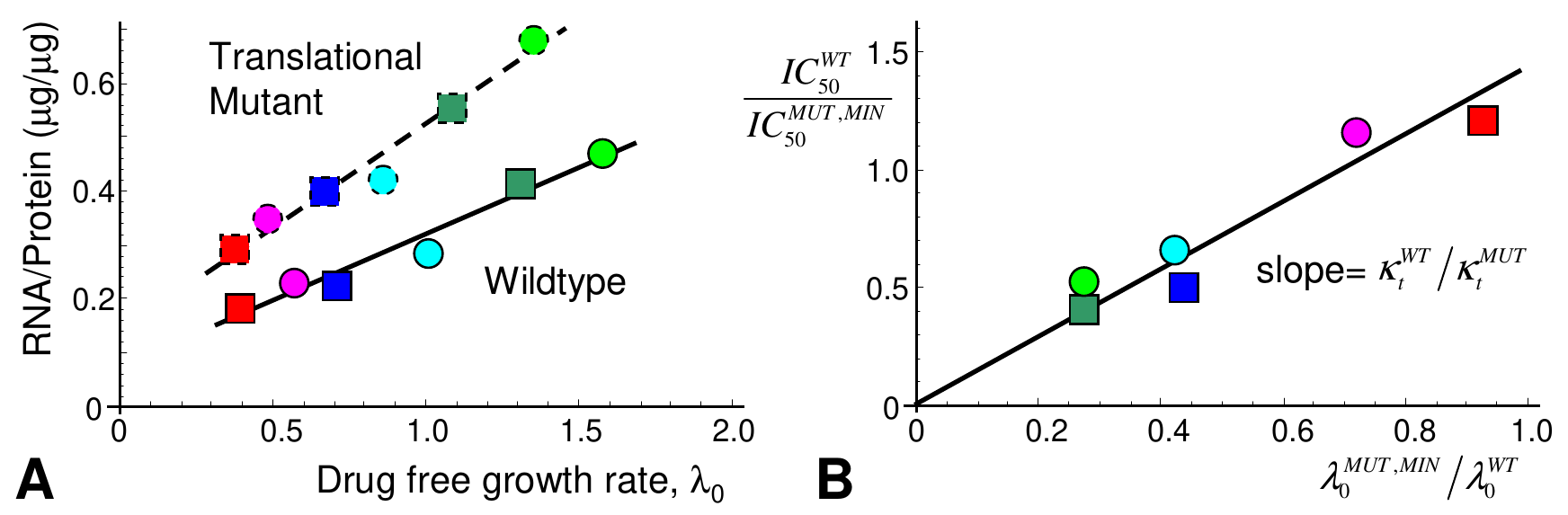}
\end{center}
\caption{The translation mutant shows growth-dependent susceptibility to tetracycline in quantitative agreement with the model predictions. {\bf{A}}: The mutant shows a reduced translational capacity compared to the wildtype strain. Translational capacity is given as the inverse slope of a plot of the RNA/protein ratio versus drug-free growth rate $\lambda_0$ (in h$^{-1}$) \cite{scott_science}. The data for the mutant are from this study (dashed line); wildtype data are taken from Scott {\it et al.} \cite{scott_science} (solid line). The ratio of slopes (WT/MUT) gives the ratio of translational capacity $\kappa_t^{\rm MUT}/\kappa_t^{\rm WT}=0.65$ (Table S5). The colored symbols indicate different growth media, as in Fig. 1.  
{\bf B}: Growth-dependent susceptibility to tetracycline for the translation mutant. The model predicts that for a reversible drug such as tetracycline, IC$_{50} = $IC$_{50}^*\lambda_0^*/(2\lambda_0)$, so that  IC$_{50}^{\rm WT}$/IC$_{50}^{\rm MUT} = (\kappa_t^{\rm WT}/\kappa_t^{\rm MUT})\times(\lambda_0^{\rm MUT}/\lambda_0^{\rm WT}) = (1/0.65)\times(\lambda_0^{\rm MUT}/\lambda_0^{\rm WT})$ (since both $\lambda_0^*$ and IC$_{50}^*$ are proportional to $\sqrt{\kappa_t}$). The symbols show IC$_{50}^{\rm WT}$ measured on all 6 growth media, divided by the IC$_{50}^{\rm MUT, MIN}$ measured on glucose minimal or glycerol minimal medium as appropriate, and the drug-free growth rate of the wildtype $\lambda_{0}^{\rm WT}$ similarly rescaled with respect to the drug-free growth rate of the mutant in the corresponding minimal medium $\lambda_{0}^{\rm MUT, MIN}$. The data collapse onto a straight line with gradient (1/0.65), as indicated by the solid black line. It is important to note that the solid line is not a line-of-best-fit, but rather comes from taking the ratio of the slopes in panel {\bf A}. \\ }
\end{figure}

\noindent {\bf Mechanistic link between reversibility timescale and growth-dependent 
susceptibility}
\vspace*{1ex}

 \begin{figure}
\begin{center}
\includegraphics[width=0.6\textwidth]{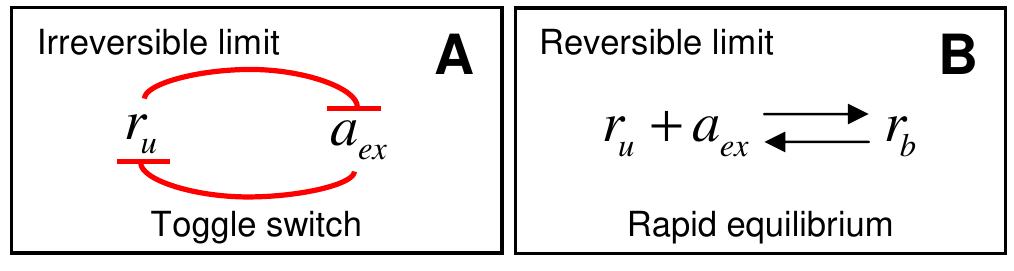}
\end{center}
\caption{Shift in the network topology in the irreversible and reversible limits. {\bf A.} In the limit that either transport or binding are irreversible (as is the case for streptomycin and kanamycin), the system exhibits a `toggle-switch' topology, leading to a steep 
inhibition curve (Eq. 11). {\bf B.} In the limit of fully-equilibrated transport and binding (as is the case for tetracycline and chloramphenicol), the model predicts more gradual inhibition (Eq. 12). \\ }
\end{figure}

\noindent Why does our model behave qualitatively differently in the limits where antibiotic transport and 
binding are irreversible (small $\lambda_0^*$) and where they are reversible (large $\lambda_0^*$)?   In the model, nutrient quality has two opposing influences on  the cell's 
ribosome content: it increases the size of the ribosome pool (solid line in Fig. 
~2B) but it also reduces the cell's capacity to increase this pool in response 
to antibiotic challenge (gradient of the dashed lines in Fig. 2B). In other 
words, fast-growing cells have a ribosome pool which is already close to maximal and have little 
capacity to increase in response to antibiotic, while slow-growing cells have a small ribosome 
pool that can be increased by a large factor in response to antibiotic. 

In the limit that either transport or binding are irreversible (small $\lambda_0^*$), antibiotic molecules that enter the cell are neutralized by binding to free 
ribosomes, such that the intracellular antibiotic concentration remains low. The model  exhibits 
a `toggle-switch' topology (Fig. 7A), in which free ribosomes ``soak up'' 
antibiotic, while antibiotic inactivates free ribosomes. If the extracellular antibiotic 
concentration $a_{\rm ex}$ is below a threshold determined by the initial (unbound) ribosome 
concentration, the cell generates ribosomes fast enough to neutralize all the antibiotic that 
enters the cell. If, however, $a_{\rm ex}$ exceeds the threshold, the cell's rate of ribosome 
generation cannot compete with the antibiotic influx and the system flips to a different 
steady-state with no free ribosomes and correspondingly no growth. Thus in the irreversible 
limit the fate of a cell is determined by a ``molecular race'' between antibiotic influx and 
ribosome production, in which the absolute number of ribosomes is decisive.
Fast-growing cells (on rich nutrient) have a larger ribosome pool and correspondingly higher  
ribosome synthesis rate, so that they are able to tolerate a higher rate of antibiotic influx 
than slow-growing cells. 

In contrast, in the limit of fully-reversible transport and binding (large $\lambda_0^*$),  the free and bound ribosome pools are in equilibrium (Fig. 
7B), and the intra- and extra-cellular antibiotic pools are also in 
equilibrium.  Increasing the antibiotic concentration shifts the   equilibrium between free and 
bound ribosome pools; the cell responds by increasing the total ribosome pool (dashed line in 
Fig. 2{B). This leads to a smoothly-varying, Langmuir-like dependence of the 
relative growth rate  $\lambda/\lambda_0$ on the extracellular antibiotic concentration 
$a_{\rm ex}$. Because  $\lambda/\lambda_0$ is determined by the {\em{relative}} sizes of the 
ribosome pool in the presence and absence of antibiotic, the half-inhibition concentration 
depends on the slope of the dashed line in  Fig. 2B. Slow-growing cells have 
more capacity to increase their ribosome pool (steeper slope of the dashed line; Fig. 2B), and as a 
consequence they are less susceptible to antibiotic than fast-growing cells. \\


\begin{center}
{\bf Discussion}
\end{center}
\vspace*{1ex}

Taken together, our results show that bacterial susceptibility to ribosome-targeting antibiotics exhibits strong growth-rate dependence, but that the nature of this dependence differs qualitatively between antibiotics (Fig. 1). For irreversibly-binding antibiotics (streptomycin and kanamycin), slower growing cells are more susceptible; whereas for reversibly-binding antibiotics (tetracycline and chloramphenicol), faster growing cells are more susceptible. This behaviour can be understood by a simple mechanistic model which shows that these contrasting effects of nutrient environment on susceptibility for different antibiotics can be explained in terms of a single parameter, the critical reversibility rate $\lambda_0^{\rm{*}}$ (Eq. 8), which characterizes the outward permeability and binding affinity of the drug.

Our model predicts a parameter-free relation for the growth-dependent susceptibility (Eq. 10), {\em i.e.}} how the IC$_{50}$ depends upon the drug-free growth rate $\lambda_0$ relative to the critical reversibility rate $\lambda_0^{\rm{*}}$. This relation is in very good agreement with the experimental data (Fig. 4).  If the pathogen drug-free growth rate $\lambda_0$ is larger than $\lambda_0^*$ the IC$_{50}$ increases with drug-free growth rate (as it does for our irreversible antibiotics streptomycin and kanamycin), so that slow-growing cells are more susceptible. In contrast, if the pathogen drug-free growth rate is smaller than $\lambda_0^*$ (as for our reversible antibiotics tetracycline and chloramphenicol) the IC$_{50}$ decreases with drug-free growth rate, so that fast-growing cells are more susceptible. Our model also predicts very different shapes for the growth-inhibition curves in these two cases; if $\lambda_0 > \lambda_0^*$ (as for our irreversible drugs), the growth-inhibition curves show a sharp drop around the IC$_{50}$, while if $\lambda_0 < \lambda_0^*$ (as for our reversible drugs), we expect smoothly varying growth-inhibition curves. Moreover,  in the reversible and irreversible limits of large and small $\lambda_0^*$, our model leads to parameter-free predictions for both growth-dependent susceptibility (Eq. 10) and growth inhibition curves (Eqs. 11 and 12),  which are confirmed by a collapse of the data points on the predicted re-scaled curves (Fig. 5). Finally, the insight provided by our analysis allows us to make sucessful predictions for how antibiotic susceptibility is modified by a mutation affecting translation rate (Fig. 6). \\

\noindent {\bf Significance of the critical reversibility rate $\lambda_0^{\rm{*}}$}
\vspace*{1ex}

\noindent A major insight arising from this study is the importance of the critical reversibility rate $\lambda_0^{\rm{*}}$ in determining susceptibility to antibiotic treatment. For a given ribosome-targeting antibiotic and pathogenic strain,  $\lambda_0^{\rm{*}}$ can be inferred from known biochemical parameters (via Eq. 8) in cases where these are known, or, alternatively, estimated by measuring inhibition curves over a range of drug-free growth rates (a task well-suited to automation \cite{bollenbach}). This critical reversibility rate 
 provides a spectrum classification of ribosome-targeting antibiotics according to their physiological effects, which, interestingly,
appears to correlate at its extremes with existing binary classification schemes, at least for the antibiotics used in this study. In particular, the irreversible antibiotics streptomycin and kanamycin are classified as bactericidal, whereas the reversible antibiotics tetracycline
and chloramphenicol are classified as bacteriostatic. This is consistent with the fact that our model predicts a rapidly-vanishing growth rate beyond the IC$_{50}$ for irreversible antibiotics ({\em i.e.} those with small values of $\lambda_0^*$).
  Our classification on the basis of $\lambda_0^* $ also correlates with the fact that  streptomycin and kanamycin are known to transiently induce expression of proteins associated with heat-shock in {\it E. coli}, whereas tetracycline and chloramphenicol induce expression of proteins associated with cold-shock~\cite{vanbogelen}. It remains to be seen whether these
responses are triggered directly by the antibiotic or are associated more generally with physiological changes occurring in the organism.\\

\noindent{\bf Coupling of cell physiology and antibiotic mode-of-action}
\vspace*{1ex}

\noindent  In a wider context, bacterial growth rate is an important factor controlling gene expression and regulation~\cite{Hwa_CR,Klumpp_Cell}, imposing strong constraints on the allocation of cellular resources. These constraints lead to intrinsic growth-rate dependence in the macromolecular composition of the cell~\cite{schaechter_rib-growthrate-rel,scott_science}. Consequently, it is to be expected (and in some cases it is known~\cite{cozens,tuomanen,millar,koch}) that antibiotic susceptibility likewise exhibits growth-rate dependence for those drugs targeting key cellular resources such as the ribosome, RNA polymerase, DNA gyrase, and cell wall biosynthetic machinery. Our results show that for ribosome-targeting antibiotics, complex growth-rate dependent susceptibility can arise from the interplay between molecular mechanism (antibiotic transport and binding) and cellular physiology (growth-dependent constraints on ribosome concentration and synthesis rate). Interestingly, our work shows that knowledge of the growth-rate dependence of the target (here, the ribosome) is not sufficient to predict the growth-rate dependence of the antibiotic susceptibility -- in fact, the nature of this dependence differs qualitatively among antibiotics despite their common target (Fig. 1). Nonetheless, contrasting patterns of growth rate-dependent susceptibility can be explained quantitatively by combining mechanistic details of antibiotic mode-of-action with empirically determined physiological constraints.

At higher concentrations than those considered here ($\sim 10 \times$ IC$_{50}$),  other 
mechanisms have been implicated in the inhibition of bacterial growth by ribosome-targeting 
antibiotics. These include  changes in the transmembrane proton-motive force, membrane 
permeabilization by  misfolded protein \cite{review_streptom_irr}, induction of a heat-shock 
response~\cite{You_bistable}, and, on longer time scales, oxidative stress which increases 
mutation rate and accelerates the emergence of resistance~\cite{Collins_ROS}. A complete picture of antibiotic action will require integration of specific response 
mechanisms, such as these, with general constraints imposed by pathogen growth, although the simple model presented here appears to capture the majority of the growth-dependent susceptibility to the ribosome-targeting antibiotics tested. Applying a similar approach to other classes of antibiotics or chemotherapeutic agents should provide a clearer picture of {\em{in vivo}} drug action.\\

\noindent{\bf Clinical and evolutionary perspectives}
\vspace*{1ex}

\noindent From a clinical perspective, the strong positive correlation of the IC$_{50}$ with drug-free 
growth rate that we observe for our irreversibly-binding antibiotics suggests that the efficacy of 
treatment could be improved by modulating the bacterial growth rate using a metabolic inhibitor 
-- echoing recent developments in understanding the role of nutrient environment in overcoming 
persistent infections~\cite{Collins_persister}. The threshold-like transition in the inhibition 
curve for irreversibly-binding antibiotics can, however, greatly facilitate acquisition of resistance, 
especially in the presence of steep spatial gradients of 
antibiotic~\cite{gradients_hermsen,deris,austin_science}, providing yet another caution against 
their improvident use~\cite{pankey}. More broadly, it is becoming clear that understanding and 
manipulating pathogen physiology plays a major role in improving strategies for the eradication 
of infection. Although both drug action and pathogen metabolism are mechanistically complex, the 
interplay between  molecular interactions and whole-cell physiology can nevertheless be 
understood quantitatively using simple rules. \\

\begin{center}
{\bf Methods}
\end{center}
 
\noindent {\em{Antibiotics}}\\
\noindent Antibiotics were obtained from Fisher Scientific: Streptomycin sulfate  
(BP910-50), Kanamycin Sulfate (BP906-5), Tetracycline hydrochloride (BP912-100) and  
Chloramphenicol (BP904-100). Stock solutions were prepared weekly, and stored at  
$4\,^{\circ}\mathrm{C}$. To avoid degradation of the antibiotics (particularly tetracycline),  
cultures were grown no longer than 6 hours before transfer to medium containing fresh antibiotic  
and all experiments were performed in light-insulated shakers.\\

\noindent {\em{Growth media}}\\
\noindent The growth media is potassium morpholinopropane sulfonate (MOPS) buffered, and is a modification  
of Neidhardt supplemented MOPS defined media~\cite{Neidhardt_Media} obtained from Teknova  
(M2101). Carbon sources used were glycerol (0.2$\%$ v/v) and glucose (0.2$\%$  
w/v). Intermediate growth rates were obtained by supplementing glycerol and glucose minimal  
media with casamino acids (0.2$\%$ w/v). The most rapid growth rates were obtained by  
supplementing the media with nucleotides (Teknova, M2103) and all amino acids  
(Teknova, M2104).\\

\noindent {\em{Strains and growth conditions}}\\
\noindent {\it Escherichia coli} K12 strain MG1655 was used in this study. Seed cultures were grown in
LB medium (Bio Basic), and used to inoculate pre-cultures in appropriate growth media without  
antibiotics. After over-night growth, pre-cultures were diluted ($500-1000 \times$) to fresh  
media and allowed to resume exponential growth for at least three generations before being  
diluted into media containing antibiotics. Cells were adapted to exponential growth in  
antibiotics and grown in adapted growth for four generations before growth rate measurements  
were taken. Cells were grown in 3 mL of culture media at $37\,^{\circ}\mathrm{C}$ in 20 mm test 
tubes, shaken in a water bath (MaxQ 7000, Thermo-Fisher) at 250 rpm. Growth rate was monitored 
by measuring $OD_{600}$ on a Biomate 3S spectrophotometer (Thermo-Fisher) over time, with cell 
viability corroborated by plating. The translational mutant strain appearing in Fig. 6 is a \emph{rpsL} point mutation that confers pseudo-dependence on streptomycin \cite{ruusala,scott_science}, moved from strain GQ9 \cite{scott_science} (also known as CH349 or UK317 \cite{ruusala}) into our wild type background via P1 transduction. \\

\noindent {\em{Protein and RNA extraction}}\\
\noindent Total protein was determined using a modified Lowry method (Sigma, TP0300) \cite{lowry, peterson}, with bovine serum albumin as a standard. RNA quantification was done via cold perchloric acid precipitation \cite{benthin}.\\

\noindent {\em{Data fits}}\\
\noindent Estimates for the critical parameter combinations $\lambda_0^{\rm{*}}$ and IC$_{50}^{\rm{*}}$ 
were obtained by fitting the experimental growth inhibition curves $\lambda(a_{\rm ex})$ to the 
solution of the cubic equation, Eq. 7. These fits were carried out using Powell's method 
\cite{numrec}.\\

\begin{center}
{\bf Acknowledgments}
\end{center}

\noindent We thank  L. Ciandrini and M. C. Romano for 
discussions and B. Waclaw, M. E. Cates, P. B. Warren, P. Swain, S. Klumpp, T. Bollenbach and R. Beardmore for 
comments on the manuscript. This work was partially supported by EPSRC under grant  
EP/J007404/1. PG was funded by a DAAD postdoc fellowship and a DFG research fellowship. RJA was 
supported by a Royal Society University Research Fellowship. MS was supported by a Discovery 
grant through the Natural Sciences and
Engineering Research Council of Canada.\\

\begin{center}
{\bf Author contributions}
\end{center}

\noindent P.G. and M.S. contributed equally to this work. M.S. performed the experiments. All authors contributed extensively to the experimental and model design, analysis of data and writing of the paper. \\

\begin{center}
{\bf Additional information}
\end{center}

\noindent The authors declare no competing financial interests.\\

\begin{center}
{\bf{Figure legends}}
\end{center}

\noindent{\bf{Figure 1}}\\
\noindent Antibiotic susceptibility depends on nutrient quality for four ribosome-targeting 
antibiotics: irreversibly-binding antibiotics streptomycin ({\bf A} \& {\bf B}) and kanamycin ({\bf C} 
\& {\bf D}), and reversibly-binding antibiotics tetracycline ({\bf E} \& {\bf F}) and  
chloramphenicol ({\bf G} \& {\bf H}). The left panels show the growth rate $\lambda$ of  
{\em{E. coli}} MG1655 relative to the drug-free growth rate $\lambda_0$, as a function of the 
antibiotic concentration.  Growth inhibition data are shown for media with glycerol as the 
carbon source. The arrows indicate increasing drug-free growth rate $\lambda_0$. The right 
panels show the half-inhibition concentration IC$_{50}$ as a function of the drug-free growth 
rate $\lambda_0$. Carbon sources are denoted by symbol: glucose (circles), glycerol (squares), 
and error bars denote the standard deviation among repeated measurements (Tables S2 and S3). Media are variants of Neidhardt's MOPS buffered medium \cite{Neidhardt_Media}; see Methods for details.\\

\noindent{\bf{Figure 2}}\\
\noindent Schematic view of the model and its dynamics. {\bf A.}  The model is focused on three 
state variables: the intracellular concentration of antibiotic $a$, the concentration 
$r_{\rm{u}}$ of ribosomes unbound by antibiotic and the concentration $r_{\mathrm{b}}$ of 
antibiotic-bound ribosomes. Two mechanisms drive the dynamics: 1. \emph{Transport} across the 
cell membrane and 2. \emph{Binding} of ribosomes and antibiotic.  {\bf B.} Constraints arising 
from empirical relations between ribosome content and growth rate. Scott {\it et al.} 
{\cite{scott_science}} measured total ribosome content as a function of growth rate. When growth 
rate is varied by nutrient composition, in the absence of antibiotics, ribosome content 
$r_{\rm{u}}$ correlates positively with growth rate $\lambda$, increasing linearly from a 
minimum concentration of inactive ribosomes $r_\mathrm{min}$ (solid line). When growth rate is 
decreased by imposing translational inhibition, total ribosome content 
$r_\mathrm{tot}=r_\mathrm{u}+r_\mathrm{b}$ increases, reaching  a maximum  $r_\mathrm{max}$ as 
growth rate decreases to zero (dashed lines). Note that Scott {\em{et al.}} measured ribosome 
mass fraction; here we translate these to concentrations (see Supplementary Information, Fig. S1).  \\

\noindent{\bf{Figure 3}}\\
\noindent Model fits to growth inhibition curve data. The parameters $\lambda_0^{\rm{*}}$ and  
${\rm IC}_{50}^*$ are obtained by numerical fitting of the solution of the cubic equation, Eq. 
7, to our experimental growth inhibition curves. Data sets for different drug-free growth rates ({\em{i.e.}} the different curves in each panel) were fitted simultaneously with the same values of  $\lambda_0^{\rm{*}}$ and  
${\rm IC}_{50}^*$, but separate fits were obtained for glycerol-based 
and glucose-based media. Here we show the resulting fits for  glycerol-based media (symbols as in Fig.~1. of the main text). The parameters obtained by this 
procedure are: Streptomycin, glycerol: $\lambda_0^{\rm{*}}=0.31$h$^{-1}$, 
${\rm IC}_{50}^*=0.19\mu$g/ml,  streptomycin, glucose: $\lambda_0^{\rm{*}}=0.57$h$^{-1}$, 
${\rm IC}_{50}^*=0.36\mu$g/ml, kanamycin, glycerol: $\lambda_0^{\rm{*}}=0.17$h$^{-1}$, 
${\rm IC}_{50}^*=0.05\mu$g/ml,  kanamycin, glucose: $\lambda_0^{\rm{*}}=0.47$h$^{-1}$, 
${\rm IC}_{50}^*=0.26\mu$g/ml, tetracycline, glycerol: $\lambda_0^{\rm{*}}=5.2$h$^{-1}$, 
${\rm IC}_{50}^*=0.23\mu$M,  tetracycline, glucose: $\lambda_0^{\rm{*}}=6.3$h$^{-1}$, 
${\rm IC}_{50}^*=0.36\mu$M, chloramphenicol, glycerol: $\lambda_0^{\rm{*}}=1.8$h$^{-1}$, 
${\rm IC}_{50}^*=2.5\mu$M,  chloramphenicol, glucose: $\lambda_0^{\rm{*}}=1.5$h$^{-1}$, 
${\rm IC}_{50}^*=4.1\mu$M. These values of $\lambda_0^*$ and ${\rm IC}_{50}^*$ are  compared to 
literature data in Table S4. Similar results are obtained if we instead fit our data directly to the predicted universal relation for IC$_{50}(\lambda_0)$ (Eq. 10); see Supplementary Information and Fig. S3.\\

\noindent{\bf{Figure 4}}\\
\noindent  Universal growth-dependent susceptibility curve. Data from the right panels of Fig. 1 
are rescaled by  $\lambda_0^{\rm{*}}$ and IC$_{50}^{\rm *}$, obtained by fitting our growth 
inhibition data (Fig.~3). The black line shows the model prediction for the universal curve, 
Eq.~10. \\

\noindent{\bf{Figure 5}}\\
\noindent Growth inhibition curves for our bactericidal and bacteriostatic drugs collapse onto 
two qualitatively different limiting forms as predicted by the model. {\bf{A}}: Data for the 
bactericidal antibiotics streptomycin (closed symbols) and kanamycin (open symbols) collapse 
onto $\lambda/\lambda_0 = (1/2)[1 + \sqrt{1 - a_{\rm ex}/{\mathrm{IC}}_{50}}]$ (black line). 
{\bf B}: Data for the bacteriostatic antibiotics tetracycline (closed symbols) and 
chloramphenicol (open symbols) collapse onto 
$\lambda/\lambda_0=1/[1+a_{\mathrm{ex}}/{\mathrm{IC}}_{50}]$ (black line).\\

\noindent{\bf{Figure 6}}\\
\noindent The translation mutant shows growth-dependent susceptibility to tetracycline in quantitative agreement with the model predictions. {\bf{A}}: The mutant shows a reduced translational capacity compared to the wildtype strain. Translational capacity is given as the inverse slope of a plot of the RNA/protein ratio versus drug-free growth rate $\lambda_0$ \cite{scott_science}. The data for the mutant are from this study (dashed line); wildtype data are taken from Scott {\it et al.} \cite{scott_science} (solid line). The ratio of slopes (WT/MUT) gives the ratio of translational capacity $\kappa_t^{\rm MUT}/\kappa_t^{\rm WT}=0.65$ (Table S5). The colored symbols indicate different growth media, as in Fig. 1.  
{\bf B}: Growth-dependent susceptibility to tetracycline for the translation mutant. The model predicts that for a reversible drug such as tetracycline, IC$_{50} = $IC$_{50}^*\lambda_0^*/(2\lambda_0)$, so that  IC$_{50}^{\rm WT}$/IC$_{50}^{\rm MUT} = (\kappa_t^{\rm WT}/\kappa_t^{\rm MUT})\times(\lambda_0^{\rm MUT}/\lambda_0^{\rm WT}) = (1/0.65)\times(\lambda_0^{\rm MUT}/\lambda_0^{\rm WT})$ (since both $\lambda_0^*$ and IC$_{50}^*$ are proportional to $\sqrt{\kappa_t}$). The symbols show IC$_{50}^{\rm WT}$ measured on all 6 growth media, divided by the IC$_{50}^{\rm MUT, MIN}$ measured on glucose minimal or glycerol minimal medium as appropriate, and the drug-free growth rate of the wildtype $\lambda_{0}^{\rm WT}$ similarly rescaled with respect to the drug-free growth rate of the mutant in the corresponding minimal medium $\lambda_{0}^{\rm MUT, MIN}$. The data collapse onto a straight line with gradient (1/0.65), as indicated by the solid black line. It is important to note that the solid line is not a line-of-best-fit, but rather comes from taking the ratio of the slopes in panel {\bf A}. \\ 

\noindent{\bf{Figure 7}}\\
\noindent Shift in the network topology in the irreversible and reversible limits. {\bf A.} In the limit that either transport or binding are irreversible (as is the case for streptomycin and kanamycin), the system exhibits a `toggle-switch' topology, leading to a steep 
inhibition curve (Eq. 11). {\bf B.} In the limit of fully-equilibrated transport and binding (as is the case for tetracycline and chloramphenicol), the model predicts more gradual inhibition (Eq. 12).

\bibliography{library_MS}




\end{document}